\documentclass[11pt]{article}
\usepackage{fullpage,epsf,amssymb,amsfonts,amsmath,latexsym, graphicx,array,extarrows,mathrsfs,mathtools,amsthm,mathrsfs}
\usepackage[normalem]{ulem}
\usepackage{authblk}

\newcommand{\overbar}[1]{\mkern 1.5mu\overline{\mkern-1.5mu#1\mkern-1.5mu}\mkern 1.5mu}

\let\oldsqrt\sqrt
\def\sqrt{\mathpalette\DHLhksqrt}
\def\DHLhksqrt#1#2{%
\setbox0=\hbox{$#1\oldsqrt{#2\,}$}\dimen0=\ht0
\advance\dimen0-0.2\ht0
\setbox2=\hbox{\vrule height\ht0 depth -\dimen0}
{\box0\lower0.4pt\box2}}

\numberwithin{equation}{section}

\title{\bf{Rigorous constraints on the matrix elements of the energy-momentum tensor}}
\author[1]{Peter Lowdon\thanks{lowdon@slac.stanford.edu}}
\author[1]{Kelly Yu-Ju Chiu\thanks{yujuchiu@stanford.edu}}
\author[1]{Stanley J. Brodsky\thanks{sjbth@slac.stanford.edu}}

\affil[1]{\small{\textit{SLAC National Accelerator Laboratory, Stanford University, 2575 Sand Hill Rd, CA 94025, USA}}}

\date{}
\begin{document}
\begin{flushright} SLAC-PUB-17111  \end{flushright}
\vspace{15mm} 
{\let\newpage\relax\maketitle}
\setcounter{page}{1}
\pagestyle{plain}

\abstract 

\noindent 
The structure of the matrix elements of the energy-momentum tensor play an important role in determining the properties of the form factors $A(q^{2})$, $B(q^{2})$ and $C(q^{2})$ which appear in the Lorentz covariant decomposition of the matrix elements. In this paper we apply a rigorous frame-independent distributional-matching approach to the matrix elements of the Poincar\'{e} generators in order to derive constraints on these form factors as $q \rightarrow 0$. In contrast to the literature, we explicitly demonstrate that the vanishing of the anomalous gravitomagnetic moment $B(0)$ and the condition $A(0)=1$ are independent of one another, and that these constraints are not related to the specific properties or conservation of the individual Poincar\'{e} generators themselves, but are in fact a consequence of the physical on-shell requirement of the states in the matrix elements and the manner in which these states transform under Poincar\'{e} transformations.

\newpage

\section{Introduction  \label{intro}}

The matrix elements of the energy-momentum tensor $T^{\mu\nu}$ are important measures of the non-perturbative structure of quantum field theories (QFTs). In particular, the form factors associated with these matrix elements have played a central role in the discussion of the spin structure of hadrons. By decomposing the angular momentum operator $J^{i}$ between hadronic spin states, sum rules are derived which attempt to connect the total angular momentum of the hadron with the spin and orbital angular momentum of its constituents~\cite{Jaffe_Manohar90, Ji_Tang_Hoodbhoy96, Ji97, Ji98, Shore_White00, Bakker_Leader_Trueman04, Leader_Lorce14,Wakamatsu14, Lowdon14}. In doing so, the form factors at zero momentum become related to one another, and are interpreted as angular momentum observables. However, it is well known that the derivation of these sum rules are beset with technical difficulties, such as the construction of well-defined normalisable hadronic states, the handling of boundary terms, and the consistent definition of the Poincar\'{e} charges~\cite{Bakker_Leader_Trueman04}. \\ 

\noindent
Axiomatic approaches to QFT provide an analytic framework from which one can analyse operator matrix elements and thus rigorously address the issues surrounding the derivation of these sum rules and their effect on the properties of the form factors. These approaches involve constructing a QFT via the definition of a series of physically motivated axioms~\cite{Streater_Wightman64,Haag96,Nakanishi_Ojima90,Bogolubov_Logunov_Oksak90,Strocchi13} such as locality and relativistic covariance of the fields. Perhaps the most significant axiom is that quantised fields are operator-valued distributions, not functions. An operator-valued distribution $\varphi$ is a continuous linear functional which maps (test) functions $f \in \mathcal{T}$ to operators $\varphi[f]$ that act on the space of states $\mathcal{H}$. In these formulations of QFT the space of test functions $\mathcal{T}$ is chosen to be $\mathcal{S}(\mathbb{R}^{1,3})$, the space of \textit{Schwartz functions}\footnote{Schwartz functions are functions of \textit{rapid decrease}~\cite{Strichartz94}.} defined on Minkowski spacetime $\mathbb{R}^{1,3}$, and the fields $\varphi$ are so-called \textit{tempered distributions}. The operator $\varphi[f]$ has an integral representation: $\varphi[f]=  \int d^{4}x \, \varphi(x)f(x)$, which gives meaning to the $x$-dependent field expression $\varphi(x)$. Although this representation is often convenient to use in calculations it should be treated carefully, since in general $\varphi(x)$ need not be continuous (e.g. the Dirac delta $\delta(x)$). From a physical perspective this corresponds to the fact that $\varphi(x)$ is not a well-defined operator, since $\varphi(x)$ would represent the performance of a measurement at a single spacetime point $x$, and this would require an infinite amount of energy~\cite{Haag96}. \\

\noindent
Another important property of distributions, which will play a central role in the calculations in this paper, is the definition of differentiation. Given a distribution $\varphi$ and test function $f$, one defines the derivative $\varphi'$ of the distribution by $\varphi'[f] := -\varphi[f']$. Since the derivative of a test function is also a test function, this definition implies that the derivative of a distribution always exists, in contrast to functions. In this sense distributions represent a generalisation of the space of functions\footnote{This explains why distributions are also often referred to as \textit{generalised functions}.}. In this paper we will demonstrate that by taking these distributional subtleties into account, one can avoid the potential inconsistencies that arise in the context of the spin sum rules and also shed new light on the behaviour of the form factors of the energy-momentum tensor.  \\

\noindent
The rest of the paper is structured as follows: first we outline the general analytic properties of the matrix elements and associated form factors of the energy-momentum tensor; using these results we then develop a novel method to constrain these form factors, and subsequently apply this method to the matrix elements of both the energy-momentum $P^{\mu}$ and angular momentum $J^{i}$ operators. Finally, we conclude by summarising our key findings.

\section{The form factors of the energy-momentum tensor   \label{T_form}}

In order to understand the structure of the matrix elements involving the energy-momentum tensor, it is important to first outline how definite momentum eigenstates are defined in axiomatic formulations of QFT. Unlike ordinary states, $|p \rangle$ is not normalisable\footnote{States of this form are often referred to as \textit{improper states}.} and therefore cannot be an element of the space of states $\mathcal{H}$. Intuitively this makes sense, since quantum uncertainty would imply that $|p \rangle$ is completely delocalised. This feature is encoded by the requirement that quantised fields are operator-valued distributions, and hence it follows that definite momentum eigenstates must be distribution-valued states~\cite{Bogolubov_Logunov_Oksak90}. In order to create a normalisable state $|\Psi^{g}_{M}\rangle$ one must therefore smear $|p\rangle$ with a test function $g \in \mathcal{S}(\mathbb{R}^{1,3})$. Moreover, to guarantee that this state is physical, and hence on shell, one requires that the test function $g$ is non-vanishing only on the upper hyperboloid $\Gamma_{M}^{+}=\{p^{2}=M^{2}, p^{0}>0 \}$, where $M$ is the mass of the physical state. Defining $\delta_{M}^{(+)}(p) := 2\pi \theta(p^{0})\delta(p^{2}-M^{2})$, this state can then be written in the form
\begin{align}
|\Psi_{M}^{g}\rangle = \int \frac{d^{4}p}{(2\pi)^{4}} \, \delta_{M}^{(+)}(p) g(p)|p  \rangle = \int \frac{d^{3}p}{(2\pi)^{3}2p^{0}}g(p)\big|_{\Gamma_{M}^{+}}\,|p \rangle,
\end{align}
where $p^{0} = \sqrt{\mathbf{p}^{2}+M^{2}}$. From the second equality it is clear that $g(p)\big|_{\Gamma_{M}^{+}}$ is actually the QFT definition of a wavepacket. For the purpose of the calculations in this paper we will consider physical spin-$\frac{1}{2}$ hadronic momentum eigenstates $|p; m; M \rangle:= \delta_{M}^{(+)}(p)|p;m\rangle$ with mass $M$ and rest frame spin projection $m \in \left\{\tfrac{1}{2},-\tfrac{1}{2}\right\}$ in the $z$-direction\footnote{We assume here that $|p ; m \rangle$ is a canonical spin state, as defined in Ref.~\cite{Bakker_Leader_Trueman04}.}. By explicitly including the factor $\delta_{M}^{(+)}(p)$ in the definition of $|p; m; M \rangle$, this ensures that these states only have support on the positive mass shell~\cite{Bogolubov_Logunov_Oksak90}. The inner product of these states is then defined in the following Lorentz-covariant manner: 
\begin{align}
\langle p';m';M|p ;m;M \rangle = (2\pi)^{4}\delta^{4}(p'-p)\delta_{M}^{(+)}(p')\delta_{m'm}
\label{norm}
\end{align} 
Eq.~(\ref{norm}) follows immediately from the standard definition of the norm\footnote{The norm of the improper states $|p ; m \rangle$ is defined by $\langle p';m'|p ;m \rangle = (2\pi)^{3}2p^{0}\delta^{3}(\mathbf{p}'-\mathbf{p})\delta_{m'm}$.} of $|p ; m \rangle$. \\

\noindent      
Now that the momentum eigenstates have been defined, one can consistently characterise the form factors associated with the energy-momentum tensor $T^{\mu\nu}$. By using the various symmetries satisfied by $T^{\mu\nu}$, the matrix elements of spin-$\frac{1}{2}$ hadronic momentum eigenstates (with mass $M$) can be written\footnote{Here we define $\sigma^{\mu\nu} = \frac{i}{2}\left[\gamma^{\mu},\gamma^{\nu}\right]$, and $a^{\{\mu }b^{\nu\}}= a^{\mu}b^{\nu}+a^{\nu}b^{\mu}$.} as follows~\cite{Ji97}:
\begin{align}
\langle p';m';M|T^{\mu\nu}(0)|p ;m;M \rangle &= \bar{u}_{m'}(p')\bigg[\frac{1}{4}\gamma^{\{\mu}(p+p')^{\nu\}}A(q^{2}) + \frac{1}{8M}(p+p')^{\{\mu}i\sigma^{\nu\}\rho}q_{\rho}B(q^{2}) \nonumber \\
& \hspace{15mm}  +\frac{1}{M}\left(q^{\mu}q^{\nu}-q^{2}g^{\mu\nu}\right)C(q^{2})    \bigg] u_{m}(p) \, \delta_{M}^{(+)}(p)\delta_{M}^{(+)}(p'),  
\label{T_decomp}
\end{align} 
where $q=p'-p$, and $u_{m}$ is the hadronic spinor. The $\delta_{M}^{(+)}$ factors reflect the fact that each state consistent with the normalisation in Eq.~(\ref{norm}) explicitly involves this factor in their definition. One can rewrite Eq.~(\ref{T_decomp}) by applying the Gordon identity, and one obtains 
\begin{align}
\langle p';m';M|T^{\mu\nu}(0)|p ;m;M \rangle &= \bar{u}_{m'}(p')\bigg[\frac{1}{8M}(p+p')^{\{\mu}(p+p')^{\nu\}}A(q^{2}) \nonumber \\
& \hspace{15mm} + \frac{1}{8M}(p+p')^{\{\mu}i\sigma^{\nu\}\rho}q_{\rho}\left[ A(q^{2})+B(q^{2}) \right] \nonumber \\
& \hspace{15mm}  +\frac{1}{M}\left(q^{\mu}q^{\nu}-q^{2}g^{\mu\nu}\right)C(q^{2})    \bigg] u_{m}(p) \, \delta_{M}^{(+)}(p)\delta_{M}^{(+)}(p').
\label{T_decomp2}
\end{align}
An important consequence of the distributional nature of the hadronic states $|p;m;M\rangle$ is that the form factors $A(q^{2})$, $B(q^{2})$ and $C(q^{2})$ are distributions, not functions. Although this feature can immediately be seen from Eq.~(\ref{norm}), which is clearly a distribution in the two variables $p'$ and $p$, this detail has largely been overlooked in the literature. The distributional nature of form factors implies that these objects are not in general point-wise defined~\cite{Streater_Wightman64}. Nevertheless, form factors $F(q^{2})$ are seemingly measured at specific values of $q^{2}$. In order to reconcile these perspectives one must recognise that one cannot ever physically measure a form factor at a specific value of $q^{2}$, since this would require an experiment with infinite precision. In practice, a measurement of $F(q^{2})$ at $q^{2}= Q^{2}$ is really a measurement of an averaged-out quantity $\overbar{F}(Q^{2};\Delta)$ in some small (but non-vanishing) region $\left[Q^{2}-\Delta,Q^{2}+\Delta \right]$. Theoretically, this is described by the fact that one must integrate Eq.~(\ref{T_decomp}) with test functions in the variables $p$ and $p'$ in order to yield a finite result, since the definite momentum eigenstates are not physical states in $\mathcal{H}$. The smearing in $p$ and $p'$ subsequently implies a smearing in $q$, and this smooths out the form factors. One ends up with expressions $\overbar{F}(Q^{2};\Delta) := (F \ast f_{\Delta})(Q^{2})$ which involve the convolution of the distribution $F$ with some test function $f_{\Delta}$, where $\Delta$ is related to the finite width associated with the wavepackets of the physical states. Since the convolution of a tempered distribution with a test function is always a smooth (infinitely differentiable) function~\cite{Strichartz94}, this explains why $\overbar{F}(Q^{2};\Delta)$ is always point-wise defined in $Q^{2}$. \\

\noindent
Since the form factor decomposition explicitly involves the factor $\delta_{M}^{(+)}(p)\delta_{M}^{(+)}(p')$, it follows that the distribution $\langle p';m';M|T^{\mu\nu}(0)|p ;m ;M\rangle$ has support for $(p',p) \in \Gamma_{M}^{+} \times \Gamma_{M}^{+}$, and therefore $A(q^{2})$, $B(q^{2})$ and $C(q^{2})$, as defined in Eq.~(\ref{T_decomp}), are restricted to have support for $q^{2} \leq 0$. Another thing to note about Eq.~(\ref{T_decomp}) is that it involves the product of distributions, which by contrast to the product of functions, is generally ill defined. Nevertheless, under certain conditions it is possible to define the product of distributions in a consistent manner~\cite{Bogolubov_Logunov_Oksak90}, and this is in fact the case for the product of $\delta^{4}(p'-p)$ and $\delta_{M}^{(+)}(p')$ in Eq.~(\ref{norm}). We will assume here that the distributional products in Eq.~(\ref{T_decomp}) are similarly well defined, and are therefore also commutative, associative, and satisfy the Leibniz rule for derivatives. \\

\noindent
Now that the structure and the distributional nature of the matrix elements of the energy-momentum tensor has been outlined, we will demonstrate in the following sections that one can obtain both model and frame-independent constraints on these form factors by decomposing the matrix elements $\langle p';m';M|P^{\mu}|p ;m;M \rangle$ and $\langle p';m';M|J^{i}|p ;m;M \rangle$ in terms of these form factors, and then comparing these decompositions with the expressions obtained after the explicit action of the Poincar\'{e} generators $P^{\mu}$ and $J^{i}$.

\section{The momentum matrix element   \label{P_matrix}}

The calculation of the matrix element $\langle p';m';M|P^{\mu}|p ;m;M \rangle$ requires one to first define the operator $P^{\mu}$ from $T^{\mu\nu}$. The standard definition of the energy-momentum operator is: $P^{\mu}= \int d^{3}x \, T^{0\mu}(x)$. However, this expression is ill-defined for several reasons\footnote{This issue has been emphasised before in several different  contexts~\cite{Lowdon14,Nakanishi_Ojima90,Strocchi13,Kastler_Robinson_Swieca66,Jegerlehner74}.}. First, since $T^{\mu\nu}$ is composed of quantised fields, it follows that it is also an operator-valued tempered distribution, and it therefore must necessarily be smeared with test functions in order to define a consistent operator $P^{\mu}$. Moreover, since these test functions belong to the space $\mathcal{S}(\mathbb{R}^{1,3})$, the integral in the definition must be performed over both space and time. In order to solve this problem, one can define the energy-momentum operator as follows~\cite{Kastler_Robinson_Swieca66,Jegerlehner74}:
\begin{align}
P^{\mu} = \lim_{\substack{d \rightarrow 0 \\ R \rightarrow \infty}}\int d^{4}x \ f_{d,R}(x) T^{0\mu}(x), \label{p_charge} 
\end{align} 
where $f_{d,R}(x) := \alpha_{d}(x_{0})F_{R}(\mathbf{x}) \in \mathcal{S}(\mathbb{R}^{1,3})$, and the test functions $\alpha_{d}$, $F_{R}$ satisfy the conditions
\begin{align}
\int dx_{0} \, \alpha_{d}(x_{0}) =1 \hspace{5mm}  &\alpha_{d}(x_{0}) \xrightarrow{d \rightarrow 0} \delta(x_{0}), \label{test1} \\
F_{R}(0) = 1 \hspace{5mm} &F_{R}(\mathbf{x}) \xrightarrow{R \rightarrow \infty} 1. \label{test2}
\end{align}
Not only does this definition guarantee that $P^{\mu}$ is convergent within matrix elements, but it also ensures~\cite{Kastler_Robinson_Swieca66} that $P^{\mu}$ is independent of the specific choice of test functions used in the limit. Using the definition in Eq.~(\ref{p_charge}), one can write\footnote{Here we have used the standard result: $T^{0\mu}(x)=e^{iP\cdot x}T^{0\mu}(0)e^{-iP\cdot x}$. However, in order for this relation to make rigorous sense one must define what one means by the operator $T^{0\mu}(0)$. $T^{0\mu}(0)$ cannot literally correspond to the $x \rightarrow 0$ limit of $T^{0\mu}(x)$ because $T^{0\mu}(x)$ is a (operator-valued) distribution, and it is therefore not point-wise defined. Instead, by $T^{0\mu}(x)$ one implicitly means the limit $n\rightarrow \infty$ of the convolution $\delta_{n}^{\{x\}}\ast T^{0\mu}$, where $\delta_{n}^{\{x\}}$ are a sequence of test functions whose support tend towards $\{x\}$ when $n \rightarrow \infty$. One then has the well-defined relation: $\delta_{n}^{\{x\}}\ast T^{0\mu} \equiv e^{iP\cdot x}\,T^{0\mu}[\delta_{n}^{\{0\}}]\,e^{-iP\cdot x}$ whose limit tends towards the intuitive result. Due to the continuity of $T^{0\mu}$ it then follows that the charge $P^{\mu}_{n}$ constructed using $\delta_{n}^{\{x\}}\ast T^{0\mu}$ converges to the definition of $P^{\mu}$ in Eq.~(\ref{p_charge}) for $n \rightarrow \infty$.}        
\begin{align}
\langle p';m';M|P^{\mu}|p ;m;M \rangle &= \lim_{\substack{d \rightarrow 0 \\ R \rightarrow \infty}} \int d^{4}x \ f_{d,R}(x) e^{iq\cdot x}   \langle p';m';M|T^{0\mu}(0)|p ;m;M \rangle \nonumber \\
&= \lim_{\substack{d \rightarrow 0 \\ R \rightarrow \infty}}\widehat{f}_{d,R}(q)  \langle p';m';M|T^{0\mu}(0)|p ;m;M \rangle. \label{p_eq1} 
\end{align}
$\widehat{f}_{d,R}(q)$ is the Fourier transform of $f_{d,R}(x)$, which due to the properties of $\mathcal{S}(\mathbb{R}^{1,3})$ is also a test function. Moreover, it follows from the conditions in Eqs.~(\ref{test1}) and~(\ref{test2}) that
\begin{align}
\lim_{\substack{d \rightarrow 0 \\ R \rightarrow \infty}}\widehat{f}_{d,R}(q) = (2\pi)^{3}\delta^{3}(\mathbf{q}). 
\label{dR_lim}
\end{align}   
Using the definition of the norm in Eq.~(\ref{norm}), and the fact that $|p;m;M\rangle$ is a momentum eigenstate, one has
\begin{align}
\langle p';m';M|P^{\mu}|p ;m;M \rangle = p^{\mu}(2\pi)^{4}\delta^{4}(p'-p)\delta_{M}^{(+)}(p')\delta_{m'm}. \label{p_eq2} 
\end{align} 
Since Eqs.~(\ref{p_eq1}) and~(\ref{p_eq2}) are equivalent representations of $\langle p';m';M|P^{\mu}|p ;m;M \rangle$, one can equate these expressions and use the form factor decomposition in Eq.~(\ref{T_decomp2}) to derive distributional constraints on $A(q^{2})$, $B(q^{2})$ and $C(q^{2})$. Doing so gives 
\begin{align}
&\delta_{M}^{(+)}(p')\Bigg[\lim_{\substack{d \rightarrow 0 \\ R \rightarrow \infty}}\widehat{f}_{d,R}(q) \, \bar{u}_{m'}(p')\bigg\{\frac{1}{8M}(p+p')^{\{0}(p+p')^{\mu\}}A(q^{2}) + \frac{1}{8M}(p+p')^{\{0}i\sigma^{\mu\}\rho}q_{\rho}\left[ A(q^{2})+B(q^{2}) \right]  \nonumber \\
&\hspace{10mm} +\frac{1}{M}\left(q^{0}q^{\mu}-q^{2}g^{0\mu}\right)C(q^{2})    \bigg\} u_{m}(p) \delta_{M}^{(+)}(p) - p^{\mu}(2\pi)^{4}\delta^{4}(p'-p)\delta_{m'm} \Bigg]_{p' \in \Gamma^{+}_{M}} =0, 
\end{align}
which under the restriction that $p' \in\Gamma^{+}_{M}$ implies the equality
\begin{align}
&\lim_{\substack{d \rightarrow 0 \\ R \rightarrow \infty}}\widehat{f}_{d,R}(q)  \, \bar{u}_{m'}(p')\bigg\{\frac{1}{8M}(p+p')^{\{0}(p+p')^{\mu\}}A(q^{2}) + \frac{1}{8M}(p+p')^{\{0}i\sigma^{\mu\}\rho}q_{\rho}\left[ A(q^{2})+B(q^{2}) \right] \nonumber \\
&\hspace{15mm} +\frac{1}{M}\left(q^{0}q^{\mu}-q^{2}g^{0\mu}\right)C(q^{2})    \bigg\} u_{m}(p) \delta_{M}^{(+)}(p) = p^{\mu}(2\pi)^{4}\delta^{4}(p'-p)\delta_{m'm}.
\label{P_equal} 
\end{align}
Because the form factors depend only on the variable $q$, it is convenient to transform to the variables $\bar{p}=\tfrac{1}{2}(p'+p)$ and $q$. Under the restriction that $p' \in\Gamma^{+}_{M}$ one can write 
\begin{align}
\delta_{M}^{(+)}(p) = 2\pi \theta(\bar{p}^{0})\frac{1}{2\bar{p}^{0}}\delta\left(q^{0} - \frac{\bar{\mathbf{p}}\cdot \mathbf{q}}{\bar{p}^{0}}\right).
\label{theta}
\end{align} 
Substituting Eq.~(\ref{theta}) into Eq.~(\ref{P_equal}) then gives
\begin{align}
&\lim_{\substack{d \rightarrow 0 \\ R \rightarrow \infty}}2\pi\widehat{f}_{d,R}(q) \delta\left(q^{0} - \frac{\bar{\mathbf{p}}\cdot \mathbf{q}}{\bar{p}^{0}}\right) \, \bar{u}_{m'}\left(\bar{p}+\tfrac{1}{2}q\right)\bigg\{\frac{1}{2M}\bar{p}^{\mu}A(q^{2}) + \frac{1}{8M\bar{p}^{0}}\bar{p}^{\{0}i\sigma^{\mu\}\rho}q_{\rho}\left[ A(q^{2})+B(q^{2}) \right]  \nonumber \\
&\hspace{15mm} +\frac{1}{2M\bar{p}^{0}}\left(q^{0}q^{\mu}-q^{2}g^{0\mu}\right)C(q^{2})    \bigg\} u_{m}\left(\bar{p}-\tfrac{1}{2}q\right) = (\bar{p}^{\mu}- \tfrac{1}{2}q^{\mu})(2\pi)^{4}\delta^{4}(q)\delta_{m'm}.
\label{eq_comp}
\end{align}   
Since $q^{\mu}\delta^{4}(q) =0$, the right-hand side of Eq.~(\ref{eq_comp}) reduces to the two variable (tensor product) distribution $(2\pi)^{4}\delta_{m'm}\,\bar{p}^{\mu}\delta^{4}(q)$. Once the limit in $d$ and $R$ is taken, it follows from Eq.~(\ref{dR_lim}) that $\widehat{f}_{d,R}(q) \, \delta\left(q^{0}-\frac{\bar{\mathbf{p}}\cdot\mathbf{q}}{\bar{p}^{0}}\right)$ tends towards the distribution $(2\pi)^{3}\delta^{4}(q)$. Using the distributional identity $h(q)\delta^{4}(q) = h(0)\delta^{4}(q)$, which holds for any infinitely differentiable function $h$, the various terms on the left-hand-side of Eq.~(\ref{eq_comp}) all end up with a different dependence on $\bar{p}$. In particular, the first term reduces to $(2\pi)^{4}\delta_{m'm}\,\bar{p}^{\mu}A(q^{2})\delta^{4}(q)$. Since this is the only term which depends on $\bar{p}$ in the same manner as the right-hand side, one can equate these expressions to obtain the (distributional) constraint   
\begin{align}
A(q^{2})\delta^{4}(q) = \delta^{4}(q).
\label{P_constr1}
\end{align}
The first thing to note with Eq.~(\ref{P_constr1}) is that the left-hand side involves a product of distributions, which due to the assumptions discussed in the previous section is well defined. By representing $\delta^{4}(q)$ as a limit of test functions $\delta_{n}^{\{0\}}(q)$, Eq.~(\ref{P_constr1}) can be written
\begin{align}
\lim_{n\rightarrow \infty}\int d^{4}q \, \delta_{n}^{\{0\}}(q)A(q^{2}) = 1,
\end{align}       
which is a rigorous formulation of the well-known result $A(0)=1$. \\    

\noindent
Since each of the remaining terms on the left-hand side of Eq.~(\ref{eq_comp}) depends on $\bar{p}$ in a different manner, and this dependence is not present in the distribution on the right-hand side, it follows that each of these distributions must individually vanish. Taking into account the fact that $\delta\left(q^{0} - \frac{\bar{\mathbf{p}}\cdot \mathbf{q}}{\bar{p}^{0}}\right)$ sets $q^{0} \rightarrow \frac{\bar{\mathbf{p}}\cdot \mathbf{q}}{\bar{p}^{0}}$, one obtains the constraints
\begin{align}
q^{j}\left[ A(q^{2})+B(q^{2})\right]\delta^{4}(q) &= q^{j}B(q^{2})\delta^{4}(q)=0,    \label{P_constr2}  \\
q^{j}q^{l}C(q^{2})\delta^{4}(q) &= 0,    \label{P_constr3}
\end{align}  
where the last equality in Eq.~(\ref{P_constr2}) follows immediately from Eq.~(\ref{P_constr1}). Since both Eqs.~(\ref{P_constr2}) and~(\ref{P_constr3}) contain explicit factors of $q$ and $\delta^{4}(q)$, without knowledge of the singular behaviour of $B(q^{2})$ and $C(q^{2})$ at $q=0$ these constraints are not particularly informative.

\section{The angular momentum matrix element  \label{J_matrix}}

By continuing with the previous approach one can now calculate the constraints on the form factors $A(q^{2})$, $B(q^{2})$ and $C(q^{2})$ from the matrix element $\langle p';m';M|J^{i}|p ;m;M \rangle$. In order to do so one must define the operator $J^{i}$. For the same reasons as with the operator $P^{\mu}$, the naive expression for the angular momentum operator $J^{i}= \frac{1}{2}\epsilon^{ijk}\int d^{3}x \, \left[x^{j}T^{0k}(x) -x^{k}T^{0j}(x)\right]$ is ill defined. Nevertheless, a consistent expression can be written in a similar manner
\begin{align}
J^{i} = \frac{1}{2}\epsilon^{ijk} \lim_{\substack{d \rightarrow 0 \\ R \rightarrow \infty}}\int d^{4}x \ f_{d,R}(x) \left[ x^{j}T^{0k}(x) - x^{k}T^{0j}(x) \right], \label{j_charge}
\end{align}
and hence the angular momentum matrix element takes the form
\begin{align}
\langle p';m';M|J^{i}|p ;m ;M\rangle &= \epsilon^{ijk} \lim_{\substack{d \rightarrow 0 \\ R \rightarrow \infty}} \int d^{4}x \ f_{d,R}(x) x^{j}e^{iq\cdot x} \langle p';m';M|T^{0k}(0)|p ;m ;M\rangle \nonumber  \\
&= -i\epsilon^{ijk} \lim_{\substack{d \rightarrow 0 \\ R \rightarrow \infty}} \frac{\partial \widehat{f}_{d,R}(q)}{\partial q_{j}} \langle p';m';M|T^{0k}(0)|p ;m ;M\rangle. \label{j_eq2}
\end{align}
However, since the states $|p ;m ;M\rangle$ transform non-trivially under rotations, the structure of the matrix element of $J^{i}$ is more complicated than the corresponding expression for $P^{\mu}$ in Eq.~(\ref{p_eq2}). To derive this expression one can use the fact that one-particle states of spin $s$ transform under (proper orthochronous) Lorentz transformations $\alpha \in \overbar{\mathscr{L}_{+}^{\uparrow}}$ as follows~\cite{Haag96}:
\begin{align}
U(\alpha)|p;k;M\rangle = \sum_{l}\mathcal{D}^{s}_{lk}(\alpha)|\Lambda(\alpha)p;l;M\rangle,
\label{Lorentz_tran}
\end{align} 
where $\mathcal{D}^{s}$ is the $(2s+1)$-dimensional Wigner rotation matrix, and $\Lambda(\alpha)$ is the four-vector representation of $\alpha$. \\

\noindent
For a general Lorentz transformation one has: $U(\alpha) = e^{i(\boldsymbol{\eta} \cdot \mathbf{K} -\boldsymbol{\beta} \cdot \mathbf{J})}$, where $J^{i}$ and $K^{i}$ are the angular momentum and boost operators, respectively. In particular, since we are interested in the matrix elements of $J^{i}$, one can consider the case where $\alpha= \mathcal{R}$ is a pure rotation, and hence: $U(\mathcal{R}) = e^{-i\boldsymbol{\beta} \cdot \mathbf{J}}$. Combining Eq.~(\ref{Lorentz_tran}) together with the definition of the norm in Eq.~(\ref{norm}), one obtains~\cite{Bakker_Leader_Trueman04}
\begin{align}
\langle p';m';M| J^{i}|p;m;M\rangle &= i\frac{\partial}{\partial \beta_{i}} \langle p';m';M|U(\mathcal{R})|p;m;M\rangle_{\boldsymbol{\beta}=0}   \nonumber \\ 
& = (2\pi)^{4}\delta_{M}^{(+)}(p')\left[S^{i}_{m'm} - i\delta_{m'm}\epsilon^{ijk}p^{j}\frac{\partial}{\partial p_{k}}  \right]\delta^{4}(p'-p), 
\label{j_eq1}
\end{align}
where $S^{i}_{m'm}:=i\frac{\partial}{\partial \beta_{i}} \left[\mathcal{D}^{s}_{m'm}(\mathcal{R})\right]_{\boldsymbol{\beta}=0}$ are the $(2s+1)$-dimensional spin matrices for spin-$s$~\cite{Bakker_Leader_Trueman04}. In order to compare the form factor expansion of Eq.~(\ref{j_eq2}) with Eq.~(\ref{j_eq1}), one must consider the specific case of spin-$\frac{1}{2}$ states, in which case: $S^{i}_{m'm} = \frac{1}{2}\sigma^{i}_{m'm}$, where $\sigma^{i}$ are the Pauli matrices. It should be noted that the remarkably simple expression in Eq.~(\ref{j_eq1}) is only true for canonical spin states~\cite{Bakker_Leader_Trueman04}, and is significantly more complicated for other spin states such as Wick helicity states\footnote{In particular, for on-shell spin-$\frac{1}{2}$ Wick helicity states $|p; m;M\rangle_{W}$ one has the following relation: ${}_{W}\langle p'; m';M| J^{i} |p; m;M\rangle_{W} = (2\pi)^4 \delta_{M}^{(+)}(p')  \left[ m\,\delta_{m'm}    \frac{ \left( \delta^{i1} p^{1} + \delta^{i2} p^{2}\right) |\mathbf{p}|}{(p^{1})^{2} + (p^{2})^{2}} -  i\delta_{m'm}\epsilon^{ijk} p^{j}\frac{\partial}{\partial p_{k}} \right] \delta^{4}(p'-p)$.}~\cite{Wick62}. \\

\noindent
Now we are in a position to perform the same matching procedure as in Sec.~\ref{P_matrix} for the matrix element $\langle p';m';M|J^{i}|p ;m;M \rangle$. Comparing the form factor expansion of Eq.~(\ref{j_eq2}) with Eq.~(\ref{j_eq1}), and substituting in Eq.~(\ref{T_decomp2}) gives
\begin{align}
&(2\pi)^{4}\left[\frac{1}{2}\sigma^{i}_{m'm} + i\delta_{m'm}\epsilon^{ijk}\bar{p}^{j}\frac{\partial}{\partial q_{k}}  \right]\delta^{4}(q) = \nonumber \\
& \hspace{10mm}  \lim_{\substack{d \rightarrow 0 \\ R \rightarrow \infty}} 2\pi i\epsilon^{ijk} \frac{\partial \widehat{f}_{d,R}}{\partial q_{k}} \delta\left(q^{0} - \frac{\bar{\mathbf{p}}\cdot \mathbf{q}}{\bar{p}^{0}}\right) \, \bar{u}_{m'}\left(\bar{p}+\tfrac{1}{2}q\right)\bigg\{\frac{1}{2M}\bar{p}^{j}A(q^{2}) \nonumber \\
& \hspace{20mm}+ \frac{1}{8M\bar{p}^{0}}\bar{p}^{\{0}i\sigma^{j\}\rho}q_{\rho}\left[ A(q^{2})+B(q^{2}) \right]  +\frac{1}{2M\bar{p}^{0}}\left(q^{0}q^{j}\right)C(q^{2})  \bigg\} u_{m}\left(\bar{p}-\tfrac{1}{2}q\right).
\label{eq_big}
\end{align} 
In order to simplify the right-hand side expression one can make use of the following set of relations~\cite{Bakker_Leader_Trueman04}:
\begin{align}
&\frac{\partial}{\partial q_{k}}\left\{\left[\bar{u}_{m'}\left(\bar{p}+\tfrac{1}{2}q\right)u_{m}\left(\bar{p}-\tfrac{1}{2}q\right)\right]_{q^{0}= \frac{\bar{\mathbf{p}}\cdot \mathbf{q}}{\bar{p}^{0}}}\right\}_{q = 0} = \frac{i}{(\bar{p}^{0}+M)}\epsilon^{kln}\, \bar{p}^{l} \,\sigma^{n}_{m'm}, \label{rel_1}\\
&\bar{u}_{m'}(\bar{p})\sigma^{jk}u_{m}(\bar{p}) = 2\epsilon^{jkl}\left[\bar{p}^{0}\sigma^{l}_{m'm} - \frac{\bar{p}^{l}(\bar{\mathbf{p}}\cdot \boldsymbol{\sigma}_{m'm})}{\bar{p}^{0}+M}   \right],
\label{rel_2} 
\end{align}
together with the distributional\footnote{In Eq.~(\ref{dist_id}) and throughout the rest of the paper $\partial^{k}$ signifies the distributional derivative with respect to $q_{k}$ when acting on distributions, and the partial derivative when acting on infinitely differentiable functions $h$.} identity~\cite{Strichartz94}
\begin{align}
h(q)\, \partial^{k}\delta^{4}(q) = h(0)\, \partial^{k}\delta^{4}(q) - (\partial^{k}h)(0)\, \delta^{4}(q). 
\label{dist_id}
\end{align}
Eq.~(\ref{dist_id}) follows directly from the definition of the derivative of a distribution discussed in Sec.~\ref{intro}. Intuitively, one might have expected only the first term, as is the case with $h(q)\delta^{4}(q)$, but in fact this second term cannot be neglected. By taking this distributional subtlety into account this resolves the issues surrounding the treatment of boundary terms in the spin sum rules in the literature~\cite{Ji_Tang_Hoodbhoy96,Bakker_Leader_Trueman04,Leader_Lorce14}. Applying Eqs.~(\ref{rel_1}), (\ref{rel_2}) and~(\ref{dist_id}) to Eq.~(\ref{eq_big}), one finally obtains  
\begin{align}
&(2\pi)^{4}\left[\frac{1}{2}\sigma^{i}_{m'm} + i\delta_{m'm}\epsilon^{ijk}\bar{p}^{j}\frac{\partial}{\partial q_{k}}  \right]\delta^{4}(q) = \nonumber \\
& (2\pi)\,\delta\left(q^{0} - \frac{\bar{\mathbf{p}}\cdot \mathbf{q}}{\bar{p}^{0}}\right)   \lim_{\substack{d \rightarrow 0 \\ R \rightarrow \infty}}\Bigg\{ \frac{1}{2}\sigma^{i}_{m'm} \, \widehat{f}_{d,R}(q) A(q^{2}) +  i\delta_{m'm}\epsilon^{ijk}\bar{p}^{j} \frac{\partial \widehat{f}_{d,R}}{\partial q_{k}} A(q^{2}) \nonumber \\
& + \left[\frac{\bar{p}^{0}}{2M} \sigma^{i}_{m'm} - \frac{\bar{p}^{i}(\bar{\mathbf{p}}\cdot \boldsymbol{\sigma}_{m'm})}{2M(\bar{p}^{0}+M)}\right] \widehat{f}_{d,R}(q) B(q^{2})  - \epsilon^{ijk}\,\frac{\bar{p}^{\{0}\bar{u}_{m'}\left(\bar{p}\right)\sigma^{j\}\rho}u_{m}\left(\bar{p}\right)q_{\rho}}{8M\bar{p}^{0}} \, \frac{\partial \widehat{f}_{d,R}}{\partial q_{k}}  \left[ A(q^{2})+B(q^{2}) \right] \nonumber \\
&  +i\delta_{m'm}\,\epsilon^{ijk}\,\frac{q^{0}q^{j}}{\bar{p}^{0}} \frac{\partial \widehat{f}_{d,R}}{\partial q_{k}} C(q^{2})   +\epsilon^{ijk} \left[i\,\frac{q^{j}\bar{p}^{k}}{(\bar{p}^{0})^{2}}\delta_{m'm} - \frac{\bar{p}_{l}q^{l}q^{j}(\bar{\mathbf{p}}\times \boldsymbol{\sigma}_{m'm})^{k}}{2M(\bar{p}^{0})^{2}(\bar{p}^{0}+M)}  \right] \widehat{f}_{d,R}(q) C(q^{2})   \Bigg\}.
\end{align}
Just as with the $P^{\mu}$ matrix element, one can match the distributions in the variable $\bar{p}$ on both sides. Since only the first two terms depend in the same manner on $\bar{p}$ as the left-hand side, one obtains the following constraints:
\begin{align}
&A(q^{2})\partial^{k}\delta^{4}(q) = \partial^{k}\delta^{4}(q), \label{J_constr1} \\
&A(q^{2})\delta^{4}(q) = \delta^{4}(q), \label{J_constr2} \\
&B(q^{2})\delta^{4}(q)= 0, \label{J_constr3} \\
&q^{l}\left[ A(q^{2})+B(q^{2})\right]\partial^{k}\delta^{4}(q) = 0,  \hspace{5mm} (l \neq k) \label{J_constr4} \\
&q^{j}q^{l}C(q^{2})\partial^{k}\delta^{4}(q) = 0,  \hspace{5mm} (l \neq k) \label{J_constr5} \\
&q^{j}C(q^{2})\delta^{4}(q) = 0. \label{J_constr6}
\end{align}
One can immediately see that the constraints derived in Sec.~\ref{P_matrix} are a subset of those above; Eq.~(\ref{J_constr2}) is identical to Eq.~(\ref{P_constr1}), and Eqs.~(\ref{P_constr2}) and~(\ref{P_constr3}) follow from Eqs~(\ref{J_constr3}) and~(\ref{J_constr6}) respectively. Thus the constraints imposed on the form factors by the Lorentz structure and support property of $\langle p';m';M|P^{\mu}|p ;m;M \rangle$ are entirely encoded in $\langle p';m';M|J^{i}|p ;m;M \rangle$. Combining Eqs.~(\ref{J_constr1}) and~(\ref{J_constr2}) gives
\begin{align}
\partial^{k}\delta^{4}(q) = \partial^{k}\left[A(q^{2})\delta^{4}(q) \right] = A(q^{2})\partial^{k}\delta^{4}(q) + \delta^{4}(q)\partial^{k}A(q^{2}), 
\end{align} 
and hence $A$ satisfies the constraint 
\begin{align}
\lim_{n\rightarrow \infty}\int d^{4}q \, \delta_{n}^{\{0\}}(q)\partial^{k}A(q^{2})  = 0,      
\end{align}
which formally corresponds to the condition $\partial^{k}A(0)=0$. In the case where $A(q^{2})$ is non-singular at $q=0$, this condition follows immediately from the fact that $A(q^{2})$ depends only on $q^{2}$, and hence derivatives with respect to $q$ must vanish at $q=0$. In much the same way that Eq.~(\ref{J_constr2}) implies that $A(0)=1$, Eq.~(\ref{J_constr3}) implies the well-known result $B(0)=0$, i.e. the anomalous gravitomagnetic moment vanishes~\cite{Brodsky_Hwang_Ma_Schmidt01}. Eqs.~(\ref{J_constr4})--(\ref{J_constr6}) are not particularly informative, but it's interesting to note that the constraint imposed on $C(q^{2})$ in Eq.~(\ref{J_constr6}) is identical to the constraint on $B(q^{2})$ in Eq.~(\ref{P_constr2}). This demonstrates that the matrix element constraints are not sufficient to extract the behaviour of $C(q^{2})$ near $q=0$. \\

\noindent
In analyses of the form factors $A(q^{2})$, $B(q^{2})$ and $C(q^{2})$ in the literature\footnote{See~\cite{Leader_Lorce14} and references within.} usually only the matrix elements of the operator $J^{3}$ are considered. By then imposing that the states have fixed momentum along the $z$-axis, and introducing appropriate wavepacket functions to ensure that the states are normalisable, the sum rule: $\frac{1}{2}= \frac{1}{2}\left[A(0)+B(0)\right]$ is obtained. It then follows from the constraint $A(0)=1$ that $B(0)=0$. In contrast, in our approach we study the distributional properties of the matrix elements, and therefore no choice of frame, wavepacket, operator component, or spin component $m$ is required. Since Eqs.~(\ref{J_constr2}) and~(\ref{J_constr3}) follow separately from the structure of the $J^{i}$ matrix element, this implies that the vanishing of $B(0)$ is actually \textit{independent} of the behaviour of $A(q^{2})$, in contrast to the literature. \\

\noindent
Interestingly, one can also analyse the matrix elements of the boost generator $K^{i}$ in a completely analogous manner to $J^{i}$ by using the transformation properties of the states $|p;m;M\rangle$ under pure boosts $U(\mathcal{B}) = e^{i\boldsymbol{\eta} \cdot \mathbf{K}}$. In this case for spin-$\frac{1}{2}$ states one has the relation~\cite{Bakker_Leader_Trueman04}
\begin{align}
\langle p';m';M|K^{i}|p ;m;M \rangle = (2\pi)^{4}\delta_{M}^{(+)}(p')\left[\frac{(\mathbf{p} \times \boldsymbol{\sigma}_{m'm})^{i}}{2(p^{0}+M)} + i\delta_{m'm}p^{0}\frac{\partial}{\partial p_{i}}  \right]\delta^{4}(p'-p). 
\label{k_eq1}
\end{align}
After equating this general expression with the energy-momentum form factor decomposition in Eq.~(\ref{T_decomp2}), and performing the same distributional matching procedure as for $J^{i}$, remarkably it turns out that one obtains precisely the same constraints as in Eqs.~(\ref{J_constr1})--(\ref{J_constr6}). Together, all of these findings demonstrate that the low-energy constraints imposed on the energy-momentum form factors are not related to the specific properties or conservation of the individual Poincar\'{e} generators themselves, as concluded in the literature~\cite{Jaffe_Manohar90,Ji97,Ji98,Shore_White00,Bakker_Leader_Trueman04,Leader_Lorce14,Wakamatsu14,Donoghue_Holstein_Garbrecht_Konstandin02}, but are in fact a consequence of the physical on-shell requirement of the states in the matrix elements and the manner in which these states transform under Poincar\'{e} transformations. \\

\noindent
Although the analysis in this paper has been performed in the instant form with canonical spin states, one could equally-well quantise the theory in the front form and using different types of spin states, and one would ultimately obtain the same form factor constraints\footnote{In particular, for light-front spin states in the front form one obtains the same expression as Eq.~(\ref{j_eq1}) for the $J^{3}$ matrix element~\cite{Chiu_Brodsky17}.}. Since the transformation properties and the positive mass-shell condition for physical states $|p;m;M\rangle$ are generic features of any QFT, it follows that the constraints in Eqs.~(\ref{J_constr1})--(\ref{J_constr6}) must hold for both free\footnote{For free fields: $A(q^{2}) \equiv 1$, $B(q^{2})\equiv 0$ and $C(q^{2}) \equiv 0$, which do indeed satisfy the constraints in Eqs.~(\ref{J_constr1})--(\ref{J_constr6}).} \textit{and} interacting theories. This is a similar situation as with the sum rules satisfied by the spectral densities of correlation functions~\cite{Lowdon15}. \\

\noindent
In this paper we have focused solely on the form factors associated with the matrix elements of the energy-momentum tensor. However, because the distributional-matching approach employed is model independent, one could in principle also use this approach to investigate the structure of form factors related to any other currents, such as the electromagnetic or axial form factors. Moreover, since the decomposition in Eq.~(\ref{j_eq1}) is valid for arbitrary canonical spin, one could generalise this approach to analyse the form factors associated with matrix elements of non-spin-$\frac{1}{2}$ states.

\section{Conclusions   \label{concl}}

Since the form factors associated with the matrix elements of the energy-momentum tensor, $A(q^{2})$, $B(q^{2})$ and $C(q^{2})$, encode the dynamics of the states involved in these matrix elements, analysing the behaviour of these objects is key to understanding the non-perturbative characteristics of any QFT. A feature which has received considerable interest in the literature, especially in the context of hadronic spin, is the behaviour of these form factors as $q \rightarrow 0$. In this paper we apply a novel axiomatic QFT distributional-matching approach to the matrix elements of $P^{\mu}$ and $J^{i}$ in order to derive $q \rightarrow 0$ constraints for these form factors. We find that these constraints imply the well-known results $B(0)=0$ and $A(0)=1$, but that in contrast with the consensus in the literature, these conditions are actually independent of one another. Furthermore, we also apply an identical procedure to the matrix elements of the boost generator $K^{i}$, and find that this leads to precisely the same constraints. These findings demonstrate that the constraints imposed on the energy-momentum form factors at zero momentum are not related to the specific properties or conservation of the individual Poincar\'{e} generators themselves, but are in fact a consequence of the physical on-shell requirement of the states in the matrix elements and the manner in which these states transform under Poincar\'{e} transformations.

\section*{Acknowledgements}
We thank Elliot Leader for useful discussions. This work was supported by the U.S. Department of Energy under contract DE--AC02--76SF00515. P.L is also supported by the Swiss National Science Foundation under contract P2ZHP2\_168622.

\renewcommand*{\cite}{\vspace*{-12mm}}

\end{document}